\documentclass{PoS}
\newcommand{\beq}{\begin{equation}}
\newcommand{\eeq}{\end{equation}}
\newcommand{\cA}{{\cal A}}
\newcommand{\cAb}{{\overline{\cal A}}}
\newcommand{\cF}{{\cal F}}
\newcommand{\cFb}{{\overline{\cal F}}}
\newcommand{\cD}{{\cal D}}
\newcommand{\cDb}{{\overline{\cal D}}}
\newcommand{\cQ}{{\cal Q}}
\newcommand{\cN}{{\cal N}}
\newcommand{\cU}{{\cal U}}
\newcommand{\cUb}{{\overline{\cal U}}} 
\newcommand{\KD}{{K\"{a}hler-Dirac }}
\newcommand{\Tr}{{\rm Tr\;}}

\newcommand{\bx}{{\bf x}}
\newcommand{\bmu}{{\bf \mu}}

\title{Supersymmetric lattices - a brief introduction}

\ShortTitle{Supersymmetric lattices}

\author{\speaker{Simon Catterall}%
        \thanks{Work supported in part by DOE grant DE-FG02-85ER40237 }\\
       Syracuse University\\
       E-mail: \email{smc@physics.syr.edu}}

\abstract{
Recently, new theoretical ideas have allowed the construction of lattice actions which are explicitly
invariant under one or more
supersymmetries. These theories are local and free
of doublers and in the case of Yang-Mills theories also possess
exact gauge invariance. In this talk these
ideas are reviewed with particular emphasis
being placed on ${\cal N}=4$ super Yang-Mills
theory.}

\FullConference{The XXVII International Symposium on Lattice Field Theory\\
                July 26-31, 2009\\
                Peking University, Beijing, China}

\begin{document}

\section{Introduction}

The problem of formulating supersymmetric theories on lattices has a long
history going back to the earliest days of lattice gauge theory. However,
after initial efforts failed to produce useful supersymmetric lattice
actions the topic languished for many years. Indeed a folklore
developed that supersymmetry and the lattice were
mutually incompatible. However, recently, the
problem has been re-examined using new tools and ideas such as
topological twisting, orbifold projection and deconstruction
and a class of lattice models have been constructed which
maintain one or more supersymmetries exactly at non-zero lattice spacing.

While in low dimensions there are many continuum supersymmetric
theories that can be discretized this way, in four dimensions 
there appears to a unique solution to the constraints -- $\cN=4$
super Yang-Mills. The availability of a supersymmetric lattice
construction for this theory is clearly very exciting from the
point of view of exploring the connection between gauge theories
and string/gravitational theories. But even without this connection
to string theory it is clearly
of great importance to be able to give a non-perturbative formulation
of a supersymmetric theory via a lattice path integral in the same
way that one can formally define QCD as a limit of lattice QCD as the
lattice spacing goes to zero and the box size to infinity.
From a practical point of view one can also hope that some of the
technology of lattice field theory such as strong coupling
expansions and Monte Carlo simulation can be brought to bear
on such supersymmetric theories.

In this talk I will outline some of the key ingredients that go into
these constructions, the kinds of applications
that have been considered so far and highlight the remaining difficulties.

First, let me explain why discretization of supersymmetric theories
resisted solution for so long. The central problem is
that naive discretizations of continuum supersymmetric theories break
supersymmetry completely and radiative effects lead to a profusion of
relevant supersymmetry breaking counterterms in the
renormalized lattice action. The coefficients to these counterterms must
then be carefully fine tuned as the lattice spacing is sent to zero
in order to arrive at a supersymmetric theory in the continuum
limit.
In most cases this is both unnatural and practically impossible -- particularly
if the theory contains scalar fields. Of course, one might have
expected problems -- the supersymmetry algebra is an extension of
the Poincar\'{e} algebra which is explicitly broken on the lattice. 
Specifically,
there are no infinitesimal translation generators on a discrete spacetime
so that the algebra $\{Q,\overline{Q}\}=\gamma_\mu p_\mu$ is already
broken at the classical level.
Equivalently it is a straightforward exercise to show that a naive
supersymmetry variation of a naively discretized supersymmetric theory
fails to yield zero as a consequence of the failure of the Leibniz rule
when applied to lattice difference operators\footnote{Significant work has
gone into generalizing the Leibniz rule to finite difference operators
in the context of non-commutative models using the techniques of Hopf
algebras see \cite{D'Adda_super,D'Adda_2d,D'Adda:2007ax}. This approach will not be discussed in this
talk}.

In the last five years or so this problem has
been revisited using new theoretical tools and ideas
and a set of
lattice models have been constructed which retain exactly
some of the continuum supersymmetry at non-zero lattice spacing.
The basic idea is to maintain 
a particular subalgebra of the full supersymmetry algebra
in the lattice theory. The hope is that this exact symmetry will
constrain the effective lattice action
and protect the theory from 
dangerous susy violating counterterms.

Two approaches have been pursued to produce such supersymmetric
actions; one based on ideas drawn from the field of topological field
theory  \cite{Catterall_topo,Sugino_sym1,Catterall_sig1,Catterall_n=2} and 
another pioneered by
David B. Kaplan and collaborators using ideas of orbifolding and deconstruction
\cite{Cohen:2003xe,Cohen:2003qw,Kaplan:2005ta}. 
Remarkably these two seemingly independent approaches lead to the
same lattice theories -- see 
\cite{Damgaard:2007eh,Damgaard_orb,Catterall:2007kn,Unsal:2006qp}
and the recent reviews \cite{Giedt:2009yd,Review}. 
This convergence of two seemingly completely different approaches leads
one to suspect
that the final lattice theories may represent essentially
unique solutions to the simultaneous requirements of locality, gauge invariance
and at least one exact supersymmetry. We will only have time to
discuss the approach via topological twisting in this talk.

\section{Topological twisting}

Perhaps the simplest way to understand how
this subalgebra emerges is to reformulate the target theory in
terms of "twisted fields". The basic idea of twisting goes back
to Witten in his seminal paper on topological field theory \cite{Witten:1988ze}
but actually had been anticipated in earlier work on
staggered fermions \cite{Elitzur:1982vh}.
In our context the idea is decompose the fields of the theory
in terms of representations not of the original (Euclidean)
rotational symmetry $SO_{\rm rot}(D)$ but a 
twisted rotational symmetry which is the
diagonal subgroup of this symmetry and an $SO_{\rm R}(D)$ subgroup
of the R-symmetry of the theory.
\beq
SO(D)^\prime={\rm diag}(SO_{\rm Lorentz}(D)\times SO_{\rm R}(D))
\eeq
To be explicit consider the case where the total number of
supersymmetries is $Q=2^D$.
In this case I can treat the supercharges of the twisted theory
as a $2^{D/2}\times 2^{D/2}$ matrix $q$. This matrix can
be expanded on products of gamma matrices
\beq
q=\cQ I+\cQ_\mu \gamma_\mu+\cQ_{\mu\nu}\gamma_\mu\gamma_nu+\ldots\eeq
The $2^D$ antisymmetric tensor components that arise in
this basis are the twisted
supercharges and satisfy a corresponding
supersymmetry algebra following from the original algebra
\begin{eqnarray}
\cQ^2&=&0\\
\{\cQ,\cQ_\mu\}&=&p_\mu\\
&\cdots&
\end{eqnarray}
The presence of the nilpotent scalar supercharge $\cQ$ is most important;
it is the algebra of this charge that we can hope to translate to
the lattice. The second piece of the algebra expresses the fact that
the momentum is the $\cQ$-variation of something which makes plausible
the statement that the energy-momentum tensor and hence the entire
action can be written in $\cQ$-exact form\footnote{Actually in the case
of $\cN=4$ there is an additional $\cQ$-closed term needed}.
Notice that an action written in such a $\cQ$-exact form is trivially
invariant under the scalar supersymmetry provided the latter remains
nilpotent under discretization.

The rewriting of the supercharges in terms of twisted variables can
be repeated for the fermions of the theory and yields a set
of antisymmetric tensors $(\eta,\psi_\mu,\chi_{\mu\nu},\ldots)$
which for the case of $Q=2^D$ 
matches the number of components of a real \KD field.
This repackaging of the fermions of the theory into a \KD field
is at the heart of how the discrete theory avoids fermion doubling
as was shown by Becher, Joos and Rabin in the early days of
lattice gauge theory \cite{Rabin:1981qj,Becher:1982ud}. 

It is important to recognize
that the transformation to twisted variables corresponds to a simple
change of variables in flat space -- one more suitable to
discretization. A true topological field theory only results when
the scalar charge is treated as a true BRST charge and attention is
restricted to states annihilated by this charge. In the language of
the supersymmetric parent theory such a restriction corresponds to a projection
to the vacua of the theory. It is {\it not} employed in these
lattice constructions.

\section{An example: 2D super Yang-Mills}

This theory satisfies our requirements for supersymmetric latticization;
its R-symmetry possesses an $SO(2)$ subgroup corresponding to rotations
of the its two degenerate Majorana fermions into each other. 
Explicitly 
the theory can be written in twisted form as
\beq
S=\frac{1}{g^2} \cQ
\int\Tr\left(\chi_{\mu\nu}\cF_{\mu\nu}+\eta [ \cDb_\mu,\cD_\mu ]-\frac{1}{2}\eta
d\right)\label{2daction}\eeq
The degrees of freedom are just the twisted fermions 
$(\eta,\psi_\mu,\chi_{\mu\nu})$ previously
described and a complex gauge field
$\cA_\mu$. The latter
is built from the usual gauge field and the two scalars present in
the untwisted theory $\cA_\mu=A_\mu+iB_\mu$ with corresponding
complexified field strength $\cF_{\mu\nu}$.

Notice that the original scalar fields transform as vectors under the
original R-symmetry and hence become vectors under the twisted
rotation group while the gauge fields are singlets under the
R-symmetry and so remain vectors under twisted rotations. This
structure makes possible the appearance of a complex gauge field in
the twisted theory. Notice though,that the theory is only
invariant under the usual $U(N)$ gauge symmetry and not its
complexified cousin.

The nilpotent transformations associated
with $\cQ$ are given explicitly by
\begin{eqnarray*}
\cQ\; \cA_\mu&=&\psi_\mu\nonumber\\
\cQ\; \psi_\mu&=&0\nonumber\\
\cQ\; \cAb_\mu&=&0\nonumber\\
\cQ\; \chi_{\mu\nu}&=&-\cFb_{\mu\nu}\nonumber\\
\cQ\; \eta&=&d\nonumber\\
\cQ\; d&=&0
\end{eqnarray*}

Performing the $\cQ$-variation and integrating out the auxiliary field
$d$ yields
\beq
S=\frac{1}{g^2}\int\Tr\left(-\cFb_{\mu\nu}\cF_{\mu\nu}+\frac{1}{2}[ \cDb_\mu, \cD_\mu]^2-
\chi_{\mu\nu}\cD_{\left[\mu\right.}\psi_{\left.\nu\right]}-
\eta \cDb_\mu\psi_\mu\right)\eeq

To untwist the theory and verify that indeed in flat space it just
corresponds to the usual theory one can do a further integration
by parts to produce 
\beq
S=\frac{1}{g^2}\int\Tr \left(-F^2_{\mu\nu}+2B_\mu D_\nu D_\nu B_\mu
-[B_\mu,B_\nu]^2+L_F\right)\eeq
where $F_{\mu\nu}$ is the usual Yang-Mills term.
It is now clear that the imaginary parts of the gauge fields $B_\mu$
can now be given an interpretation as
scalar fields in this parameterization. 
Similarly one can build spinors out of the
twisted fermions and write the action in the manifestly Dirac form
\beq
L_F=
\left(\begin{array}{cc}\chi_{12}&\frac{\eta}{2}\end{array}\right)
\left(\begin{array}{cc}-D_2-iB_2&D_1+iB_1\\
                        D_1-iB_1&D_2-iB_2\end{array}\right)
\left(\begin{array}{c}\psi_1\\ \psi_2\end{array}\right)
\eeq

\section{Discretization}

The prescription for discretization is somewhat natural. (Complex)
gauge fields are represented as complexified Wilson gauge links
$\cU_\mu(x)=e^\cA_\mu(x)$ living on links of a lattice which for the moment we
can think of as hypercubic.
These transform in the usual way under $U(N)$ lattice gauge transformations
\beq
\cU_\mu(x)\to G(x)\cU_\mu(x)G^\dagger(x)\eeq
Supersymmetric invariance
then implies that $\psi_\mu(x)$ live on the same links
and transform identically. 
The scalar fermion $\eta(x)$ is clearly most naturally associated with
a site and transforms accordingly
\beq \eta(x)\to G(x)\eta(x)G^\dagger(x)\eeq
The field $\chi_{\mu\nu}$ is slightly more difficult. Naturally as a 2-form
it should be associated with a plaquette. In practice we introduce diagonal
links running through the center of the plaquette and choose $\chi_{\mu\nu}$
to lie {\it with opposite orientation} along those diagonal links. This
choice of orientation will be necessary to ensure gauge
invariance.
\begin{center}\includegraphics[width=0.5\textwidth]{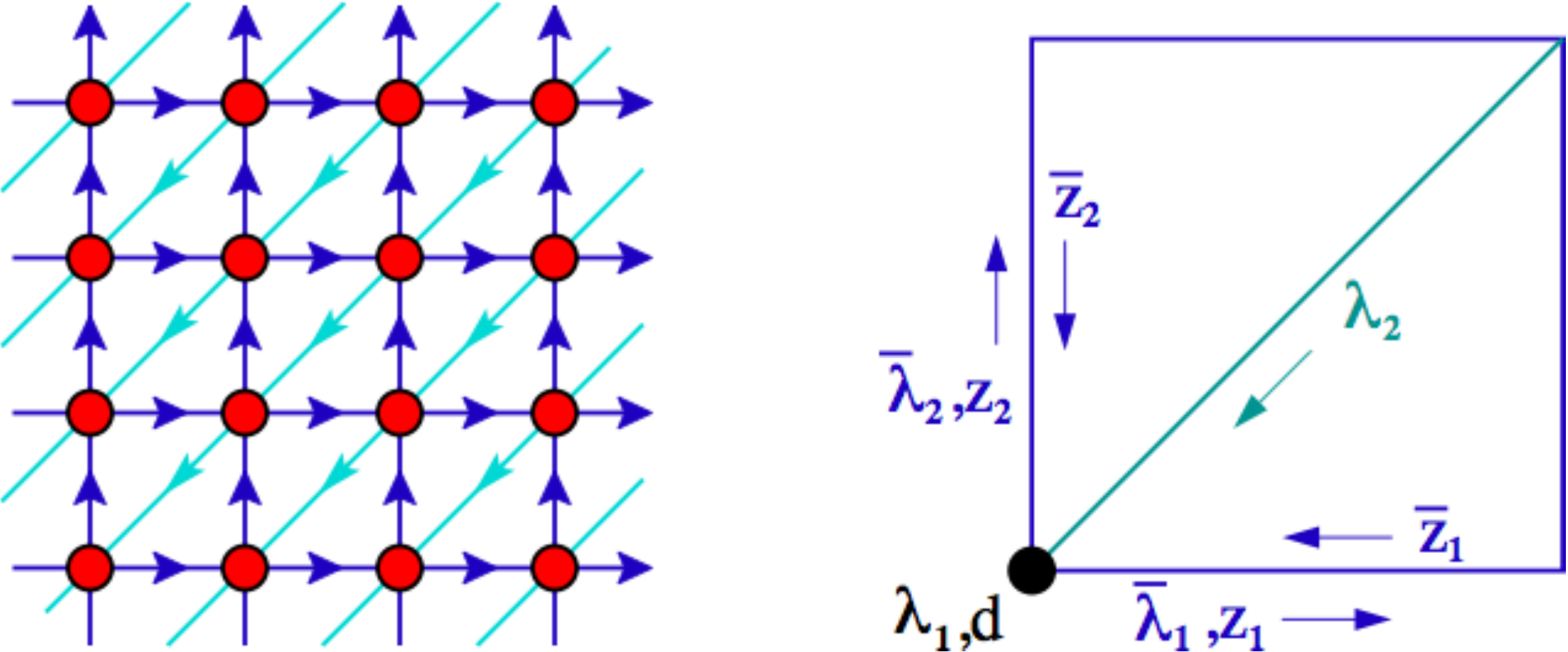}\end{center}

To complete the discretization we need to describe how continuum derivatives
are to be replaced by difference operators. A natural technology for
accomplishing this in the case of adjoint fields was developed many
years ago and yields expressions for
the derivative operator applied to arbitrary
lattice p-forms \cite{Aratyn}. 
In the case discussed here we need just two derivatives given by
the expressions
\begin{eqnarray}
\cD^{(+)}_\mu f_\nu&=&
\cU_\mu(x)f_\nu(x+\mu)-f_\nu(x)\cU_\mu(x+\nu)\\
\cDb^{(-)}_\mu f_\mu&=&f_\mu(x)\cUb_\mu(x)-\cUb_\mu(x-\mu)f_\mu(x-\mu)
\end{eqnarray} 
The lattice field strength is then given by 
the gauged forward difference $\cF_{\mu\nu}=D^{(+)}_\mu \cU_\nu$
and is automatically antisymmetric in its indices.
Furthermore it transforms like
a lattice 2-form and
yields a gauge invariant loop on the lattice when contracted
with $\chi_{\mu\nu}$.
Similarly the covariant backward difference appearing in $\cDb_\mu \cU_\mu$
transforms as a 0-form or site field and hence can be contracted with
the site field $\eta$.

This use of forward and backward difference operators guarantees that the
solutions of the theory map one-to-one with the solutions of the continuum
theory and hence fermion doubling problems are evaded.
Indeed, by introducing a lattice with half the lattice spacing one can
map this \KD fermion action into the action for staggered fermions. 
Notice that, unlike the case of QCD, there is no rooting problem in
this supersymmetric construction since the additional fermion
degeneracy is already required by the continuum theory.

\section{Twisted $\cN=4$ super Yang-Mills}

In four dimensions the constraint that the target theory possess 16
supercharges singles out a single theory for which this
construction can be undertaken -- $\cN=4$ SYM. 

The continuum twist of $\cN=4$ that is the
starting point of the twisted lattice construction was
first written down by Marcus in 1995 \cite{Marcus} although it now plays a 
important role
in the Geometric-Langlands program and is hence sometimes called
the GL-twist \cite{Kapustin:2006pk}. This four dimensional twisted theory 
is most compactly expressed
as the dimensional reduction of a five dimensional theory in which the
ten (one gauge field and six scalars) bosonic fields
are realized as the components of a complexified five dimensional
gauge field while the 16 twisted fermions naturally span one of the
two \KD fields needed in five dimensions. Remarkably, the action of
this theory contains a $\cQ$-exact piece of precisely the same
form as the two dimensional theory given in eqn.~\ref{2daction}
provided one extends the field labels to run now from
one to five. In addition the Marcus twist requires
a new $\cQ$-closed term which was not possible in the two
dimensional theory. 
\beq
S_{\rm closed}=-\frac{1}{8}\int \Tr 
\epsilon_{mnpqr}\chi_{qr}\cDb_p\chi_{mn}\label{closed}\eeq
The supersymmetric invariance of this term then relies on the
Bianchi identity $\epsilon{mnpqr}\cD_p\cF_{qr}=0$.

The lattice that emerges from examining the
moduli space of the lattice theory 
is called $A_4^*$ and is constructed
from the set of five basis vectors $v_a$ pointing out from the center of
a four dimensional equilateral simplex out to its vertices together with their
inverses $ -v_a$. It is the four dimensional analog of
the 3D bcc lattice. Complexified Wilson gauge link variables $\cU_a$
are placed on these links together with their $\cQ$-superpartners
$\psi_a$. Another 10 fermions are associated with the
diagonal links $v_a+v_b$ with $a>b$. Finally, the exact
scalar supersymmetry implies the existence of a
single fermion for every lattice site.
The lattice action corresponds to a discretization of the Marcus twist
on this $A_4^*$ lattice and can be represented as a set of traced
closed bosonic and fermionic loops. It is invariant under the exact $\cQ$ scalar
susy, lattice gauge transformations and a global permutation symmetry
$S^5$ and can be proven free of fermion doubling problems as
discussed before.

While the supersymmetric invariance of the $\cQ$-exact term is manifest
in the lattice theory it is not clear how to discretize the
continuum $\cQ$-closed term. Remarkably, it is possible to discretize eqn.~\ref{closed}
in such a way that it is indeed exactly invariant under the twisted
supersymmetry! The discrete gauge invariant form is given by
\beq
S_{\rm closed}=-\frac{1}{8}\sum_{\bx}\Tr 
\epsilon_{mnpqr}\chi_{qr}(\bx+\bmu_m+\bmu_n+\bmu_p)
\cDb^{(-)}_p\chi_{mn}(\bx+\bmu_p)\eeq
and can be seen to be supersymmetric since the lattice field
strength satisfies an exact Bianchi identity of the form
\beq
\epsilon_{mnpqr}\cD^{(+)}_p\cF_{qr}=0\eeq

\section{Prospects}
One of the key issues that still remains to be explored is the
question of how much residual fine tuning will be required to achieve
a continuum limit in which full supersymmetry is restored. This is controlled
by the flows in all relevant operators which could be induced in the
effective action as a result of quantum corrections.
We have used the exact lattice symmetries together with power
counting to enumerate the possible set of such lattice operators.

Only three terms appear and two of these correspond to renormalizations of
kinetic terms already present in the bare lattice action. There is
one additional term of the form
\beq
\eta \cU_\mu\cUb_\mu\eeq
which leads to supersymmetric mass terms for the fermions and scalars.

The question of the restoration of full
supersymmetry then rests on whether the ratios of
the coefficients to these operators
flow away from their classical values as the lattice spacing is
decreased. A one loop calculation is in progress which should
shed light on this issue. 

Beyond this issue it would be very interesting to use Monte Carlo
simulation to test the aspects of the AdS/CFT
conjecture. Parallel code has been developed to
study $\cN=4$ super Yang-Mills. At finite temperature this theory
and its dimensional reductions should be dual to a variety of
black hole solution in supergravity. 
We hope to report on the results of
these soon. One would hope that the results of such simulations could
be useful in the quest to understand how aspects of the 
quantum geometry
can be understood in terms of the dual gauge theory.

\end{document}